\documentclass{epl}
\usepackage{amsmath}
\usepackage{amsfonts}
\usepackage{amssymb}
\usepackage{mathrsfs}
\usepackage{color}
\usepackage{graphicx}

\newcommand{\bc}{\begin{center}}
\newcommand{\nno}{\nonumber}
\newcommand{\ec}{\end{center}}
\newcommand{\be}{\begin{eqnarray}}
\newcommand{\ee}{\end{eqnarray}}

\newcommand{\omits}[1]{}
\newcommand{\GR}{${\cal GR}$}
\newcommand{\dS}{$d{\cal S}$}

\definecolor{dyellow}{rgb}{1.,0.8,.0}
\definecolor{myblue}{rgb}{.1,.1,.7}
\definecolor{dcyan}{rgb}{.0,.6,.6}
\definecolor{dmagenta}{rgb}{0.6,0.0,0.6}
\definecolor{brown}{rgb}{0.6,0.2,0.}
\definecolor{darkblue}{rgb}{.0,.0,0.5}
\definecolor{darkred}{rgb}{0.75,0.0,0.0}
\definecolor{orange}{rgb}{1.,.6,.0}
\definecolor{dorange}{rgb}{0.8,.4,.0}
\definecolor{darkgreen}{rgb}{0.0,0.6,0.0}
\definecolor{purple}{rgb}{.4,.0,.4}

\newcommand{\La}{\Lambda}

\newcommand{\Om}{\Omega}

\newcommand{\dl}{\delta}

\newcommand{\ka}{\kappa}
\newcommand{\la}{\lambda}
\newcommand{\si}{\sigma}

\newcommand{\vph}{\varphi}

\newcommand{\del}{\nabla}

\title{Temperature at horizon in de Sitter spacetime}
\shorttitle{Temperature at horizon in dS spacetime}

\author{H.-Y. Guo\inst{1,2} \thanks{Email: \email{hyguo@itp.ac.cn}}
  \and
  C.-G. Huang\inst{1}
 \thanks{Email: \email{huangcg@ihep.ac.cn}}
  \and
  B. Zhou\inst{2,3} \thanks{Email: \email{zhoubihn@yahoo.com.cn}}
}
\institute{
  \inst{1} Institute of High Energy Physics, Chinese Academy of Sciences -
  P.O. Box 918-4, Beijing 100049, China \\
  \inst{2} Institute of Theoretical Physics, Chinese Academy of Sciences -
  P.O.Box 2735, Beijing 100080, China \\
  \inst{3} Department of Physics, Beijing Normal University - Beijing 100875,
  China
}
\pacs{98.80.Jk}{Mathematical and relativistic aspects of cosmology}
\pacs{11.10.Wx}{Finite-temperature field theory}
\pacs{04.70.Dy}{Quantum aspects of black holes, evaporation, thermodynamics}

\begin{document}

\maketitle

\begin{abstract}
It is found that there is no period in the imaginary Beltrami-time
of the de Sitter spacetime with Beltrami metric and that the
`surface-gravity' in view of inertial observers in de Sitter
spacetime is zero!  They show that the horizon might be at
zero temperature in de Sitter spacetime and that the
thermal property of the horizon in the de Sitter spacetime with a static metric
should be analogous to that of the Rindler horizon in Minkowski
spacetime.
\end{abstract}



\section{Introduction}

Recent observations show that our universe is dark and in
accelerated expansion \cite{90s,WMAP}.
It implies that our universe is probably asymptotically de Sitter
(\dS) with a tiny positive cosmological constant $\La$.
It is well known that \dS-universe is empty and of constant curvature
without intrinsic singularity but there is a cosmological horizon
surrounding an observer at the spatial origin.  The horizon is
naturally associated with a temperature \cite{FH-KN, GH77, GMG} and an entropy proportional to the area of the horizon \cite{GH77}.
The standard approaches in the framework of general relativity (\GR)
treat the thermodynamic aspects of the \dS-universe and black holes in
the same way \cite{GH77,GMG,review,HLW}. This leads to the puzzles on
\dS-universe \cite{Str}:  why \dS-universe is like a black hole thermodynamically and what is the statistical origin of
the cosmological horizon entropy?  A lot of
proposals have been presented to this puzzle so far, but there is
not any satisfactory explanation to them yet~\cite{JP}.

As is well known, in Minkowski spacetime the Poincar\'e-invariant field
theories are defined in inertial frames and are at zero temperature.
While in the Rindler metric which describes a special non-inertial
frame in which each observer moves at a constant 4-acceleration,
there is a horizon with Hawking temperature.  Thus, the
Hawking temperature observed by Rindler observers should be aroused
by non-inertial motions in the flat spacetime rather than gravity.
In fact, it has been pointed out that the temperature is usually
regarded as a kinetic effect, depending on the coordinate chart used by a class of observers, but not a property of the spacetime geometry in general \cite{Unruh, Deser,Pad}.

In this letter, we show that there are special coordinate systems in
\dS-spacetime, called the Beltrami coordinate systems
\cite{beltrami}, which play the role of Minkowski coordinate systems
in Minkowski spacetime and argue that the horizon in the
\dS-spacetime with a Beltrami metric should be at zero temperature
and that the relation between Beltrami and static metrics of \dS-spacetime is
analogous to that between Minkowski and Rindler metrics of flat spacetime.


\section{\dS-spacetime in static coordinates}\label{SdS}

The \dS-spacetime with cosmological constant $\Lambda$ can be viewed as a 4-d hyperboloid
\begin{equation}
\label{5sphr}
  {\cal S}_\Lambda: \eta^{}_{AB} \xi^A \xi^B =
  -R^2,\qquad R^2:=3\Lambda^{-1},
\end{equation}
embedded in a 5-d Minkowski-spacetime with
\begin{eqnarray}
\label{ds2}%
&&\upd s^2=\eta^{}_{AB} \upd\xi^A \upd\xi^B, \quad \\%
&&(\eta_{AB})={\rm diag}(1, -1, -1, -1, -1), \qquad {A, B=0,\cdots,4}.%
\end{eqnarray}
${\cal S}_\Lambda$ and the metric on it are invariant under the \dS-group
$SO(4,1)$.

The \dS-spacetime in static coordinates
\begin{eqnarray}\label{dSstatic}
 \upd s_s^2 = (1-r_s^2/R^2)\upd t_s^2 - \frac
{\upd r_s^2}{1-r_s^2/R^2}-r_s^2\upd\Om^2, \end{eqnarray}
where $\upd\Om^2$ is the standard metric on the unit sphere, can be obtained
by setting
\begin{eqnarray}\label{tperiod}
\xi^0 &=& (R^2-r_s^2)^{1/2}\sinh (t_s/R), \nno\\
\xi^1 &=& r_s \sin \theta \cos \vph, \nno \\
\xi^2 &=& r_s \sin \theta \sin \vph, \\
\xi^3 &=& r_s \cos \theta, \nno \\
\xi^4 &=&(R^2-r_s^2)^{1/2}\cosh (t_s/R). \nno
\end{eqnarray}
The above static coordinate system only covers part of the
region $\xi^0+\xi^4 > 0$, $\xi^0-\xi^4<0$, $(\xi^1)^2 + (\xi^2)^2
+ (\xi^3)^3 < R^2 \subset {\cal S}_\La$.

The null hypersurface $r_s=R$ is a horizon~\cite{GH77}, { which can
be regarded as the boundary of the static region and is at
$(\xi^0)^2=(\xi^4)^2$}.  Its cross section with a spacelike
hypersurface is a two-sphere with area $A= 4\pi R^2$.  According to
the standard approach in \GR, the surface gravity on the
horizon is defined by \cite{wald}
\begin{eqnarray} %
 \kappa_s:=\lim_{r_s \to R}(Va)= R^{-1}, %
\end{eqnarray}
where $V=(1-r_s^2/R^2)^{1/2}$ is the redshift factor and $a=(-a^\mu
a_\mu)^{1/2}$ is the magnitude of 4-acceleration of a static
observer with 4-velocity $U^\mu={ \dl^\mu_0}$ near the horizon.
The 4-acceleration $a^\mu:=(U^\nu\del_\nu U^\mu)/(U^\la U_\la)=-r_s/R^2\dl^\mu_1$.
The surface gravity has the similar explanation to the one for a
static black hole:  $Va$ is the force that must be exerted at the
origin to hold a unit test mass in place and $\kappa_s$ is the
limiting value of this force at the horizon~\cite{wald}.  The Hawking
temperature is $T_{d\cal S}={\kappa_s}/{(2\pi)}$.

\section{The Beltrami metric and its relation with the static metric }
By use of  the Beltrami coordinates\cite{SRBdS, SRBdSc,
beltrami}
\begin{eqnarray} \label{x}%
x^\mu &=& R {\xi^\mu}/{\xi^4},  \\
\xi^4 &=& \pm(\eta_{\mu \nu }{\xi ^\mu} {\xi^\nu}+R^2)^{1/2} \neq 0, \label{u4}
\end{eqnarray}%
where $\eta_{\mu \nu}={\rm diag}(1,-1,-1,-1)$, $\mu,
\nu=0,\cdots,3$, the metric (\ref{ds2}) leads to
the Beltrami form %
\begin{eqnarray} \label{B1} %
\upd s^2=\left ({\sigma^{-1}(x)}{\eta_{\mu \nu}}+ {
R^{-2}\sigma^{-2}(x)}{\eta_{\mu \la}\eta_{\nu \si}x^\la
x^\si}\right )\upd x^\mu \upd x^\nu ,%
\end{eqnarray}
\begin{equation}\label{domain} %
  \sigma(x):=\sigma(x,x)>0,
\end{equation} %
with $\sigma(a,b):=1-R^{-2} \eta_{\mu \nu}a^\mu b^\nu$.
Locally, these coordinates are similar to the inhomogeneous
coordinates in projective geometry except antipodal
identifications which cause the non-orientation problem. The
metric (\ref{B1}) only covers the region (or patch) $\xi^4 >0 $ or
$\xi^4<0$, denoted by $U_{\pm 4}$, respectively.  In order to
cover the whole $d{\cal S}$ spacetime patch by patch, one needs at least 8
patches $U_{\pm\alpha}:= \{ \xi\in{\cal S}_\Lambda |
 \xi^\alpha\gtrless 0\}, \alpha=1,\cdots, 4$.
Note that eq.~(\ref{B1}) and the inequality (\ref{domain}) are
{\it invariant}
under the factional linear transformations ($FLT$s)
\begin{equation}\label{G}
\begin{array}{l}
x^\mu\rightarrow \tilde{x}^\mu=\pm\sigma^{1/2}(a)\sigma^{-1}(a,x)(x^\nu-a^\nu)D_\nu^\mu,\\
D_\nu^\mu=L_\nu^\mu+R^{-2}\eta_{\nu\la}a^\la a^\si (\sigma(a)+\sigma^{1/2}(a))^{-1}L_\si^\mu,\\
L:=(L_\nu ^\mu) \in SO(3,1),
\end{array}\end{equation}
and {\it invariant} under the coordinate transformation in
the
intersection of two different patches, which
also takes a fractional linear form.  For instance, in $U_3$ the Beltrami coordinates are defined by
\begin{eqnarray}  
  y^{\nu'} = R {\xi^{\nu'}}/{\xi^3},\quad \textrm{where } \xi^3 > 0, \quad %
  \nu'=0, 1, 2, 4. %
\end{eqnarray} %
In $U_4\bigcap U_3$, the transition function $T_{4,3}=
{\xi^3}/{\xi^4}=x^3/R=R/{y^4}\in SO(3,1)$ and $x^i=T_{4,3}y^{i}$ and
$x^3=R^2/y^4$ is also an $FLT$.  All the $FLT$s form the group
$SO(4,1)$.

In \dS-spacetime, all geodesics satisfy linear equations in Beltrami
coordinate systems \cite{SRBdS,SRBdSc}.  They are known as {\it
straight} lines in the sense of analytic geometric approach to the
non-Euclidean geometry, first given by Beltrami long time ago for the
Lobachevsky plane \cite{beltrami}. In particular, the Beltrami-time
axis is a timelike straight world line and other three Beltrami
coordinate axes are spacelike straight lines. Conversely, if a curve
appears as a straight line in the above sense, it is a geodesic
\cite{SRBdS,SRBdSc}.  In addition, the linearity of the
equations, which the geodesics satisfy, is invariant under the
$FLT$s of $SO(4,1)$.

In physics, in a Beltrami frame on the \dS-spacetime, a free
massive particle
moves uniformly on a straight line, which is a timelike geodesic~\cite{SRBdS,SRBdSc
}:
\begin{equation}\label{imotion} %
x^\mu=v^\mu t +a^\mu, %
\end{equation} %
where $a^0 =0$, $v^0=c=1$, $a^i =x^i|_{t=0}, i=1,2,3,$ and $v^i$
are constants.
Similarly, a free light signal moves uniformly on a straight line,
which is a null geodesic, and whose trajectory obeys the
linear equation as above \cite{SRBdS,SRBdSc}. Under the $FLT$s of
$SO(4,1)$, the property of a uniform-velocity motion along a
straight line is invariant.  These uniform motions along straight
lines are referred to as {\it a kind of inertial motions in
\dS-spacetime}. In addition, two nearby inertial observers `at
rest' ($v^i=0$) in a Beltrami frame will be unchanged in the evolution
 i.e.
\begin{eqnarray} \label{Bgde}%
\frac {\upd \dl x^\mu}{\upd t} =0 \quad \mbox{and} \quad
\frac {\upd^2 \dl x^\mu}{\upd t^2} =0.\end{eqnarray}
Thus the role of the Beltrami coordinates in \dS-spacetime is
similar to that acted by the Minkowski coordinates in special 
relativity. In fact,
since the inertia properties of motions and coordinates are
invariant under $FLT$s of $SO(4,1)$, the de Sitter invariant special
relativity can be set up
\cite{SRBdS,SRBdSc}.

It is remarkable that both the static metric (\ref{dSstatic}) and
the Beltrami metric (\ref{B1}) are meaningful after Wick rotations.
The former is well known.  The latter is the 4-d Riemann sphere in
the Beltrami coordinate system \cite{beltrami,SRBdS,Pauli},
\be\label{4Bds}%
 \upd s_E^2&=&\{\delta_{\mu\nu}\sigma_E^{-1}(x)
 - R^{-2}\sigma_E^{-2}(x)\delta_{\mu \ka}x^\ka\delta_{\nu \la}x^\la \}
 \upd x^\mu \upd x^\nu,\\\label{sigma}%
&&\sigma_E(x):=\sigma_E(x,x)>0,
 \ee%
where subscript $E$ indicates the Euclidean signature version and $\sigma_E(x,x)=1+R^{-2}\delta_{ij}x^ix^j$.
Eqs.~(\ref{tperiod}) and (\ref{x}) indicate that the
`cosmic time' $t_s$ in metric (\ref{dSstatic}) and the Beltrami 
time $t$ on $U_4$ satisfy
\begin{eqnarray}\label{tperiod1}%
t|_{U_4}:=R\frac{\xi^0}{\xi^4}=R\tanh(t_s/R), \quad {\rm if} \quad \xi^4\neq 0.%
\end{eqnarray}%
It shows that there is a period in the imaginary `cosmic time'
$t_s$ and there is no period in the imaginary Beltrami time.
This is just the case: the coordinate axis of $t$ is a
straight-line and there is no period in the imaginary time $it$.
Note that the period of tangent function may be $\pi$ or $2\pi$,
depending on whether the antipodal on a unit circle is identified or not.
As mentioned before, the antipodal identification in \dS-spacetime
is excluded \cite{SRBdS,SRBdSc}. Hence, the period of the tangent
function in the Euclidean version of (\ref{tperiod1}) should be
$2\pi$.  By the finite-temperature Green's function theory, the
`Hawking temperature' $T$ is equal to the inverse of the period
$(2\pi R)$ in the imaginary time $it_s$
in static coordinates.

Does this imply, in comparison with the relation between imaginary
Minkowski time and imaginary Rindler time, that the vacuum in
\dS-spacetime in view of inertial observers should be
at zero-temperature?  One may argue that the Hawking temperature of
\dS-spacetime is determined by the surface gravity on the horizon via
the Bekenstein-Hawking relation and that the surface gravity on the
horizon is coordinate-independent. Therefore, the above analysis
does not mean that the vacuum should be at zero-temperature.  In the
following, however, we shall show that the surface gravity in
\dS-spacetime may have different explanation in view of inertial
observers in de Sitter invariant special relativity.


\section{ The horizon in \dS-spacetime with a Beltrami metric and `surface gravity'}\label{horizon}


For two events $A(a^\mu)$ and $B(b^\nu)$ in a Beltrami frame
on \dS-spacetime, the quantity
\begin{eqnarray}\label{cross}%
{\Delta}_{ R}^2(a,
b)=R^2\{\sigma^{-1}(a)\sigma^{-1}(b)\sigma^2(a,b)-1\}%
\end{eqnarray}%
is an invariant under $FLT$s of $SO(4,1)$.  The interval between a
pair of events $A$ and $B$ is said to be time-like, null or
space-like if $\Delta_{ R}^2(a, b)\gtreqless 0$,
respectively.
The light-cone with running points $X(x^\mu)$ at top $A(a^\nu)$ is given by
\begin{eqnarray} \label{nullcone} %
f=R^2 \{\sigma(a,x) \mp [\sigma(a)\sigma(x)]^{1/2}\}=0.%
 \end{eqnarray}%

It can be shown that in the Beltrami coordinates, the event horizons
in the \dS-spacetime at $(\xi^0)^2=(\xi^4)^2$ correspond to the
hypersurface $t=\pm R$ as shown in Fig. 1, which are null, of
course.
In general, an arbitrary light cone in \dS-spacetime is a
null hypersurface tangent to its  conformal boundary $\mathscr{I}^+
+ \mathscr{I}^-$: $\sigma(x)=0$. For example, when the top $A(a)$ of
a light cone tends to the point $A'$ on  $\mathscr{I}^+$,
$\sigma(a')=0$, the past light cone with top $A$ tending to $A' \in
\mathscr{I}^+$ should give rise to the future event horizon for the
observer, represented by a world line $\gamma$ which $A$ belongs to.
In fact, the horizon for the observer can be given in Beltrami
coordinates by:
\begin{eqnarray}\label{Heq}%
\lim_{a \to a'} \sigma (a,x)=0, \qquad \lim_{a\to a'}\sigma(a)=0.%
\end{eqnarray}
It is easy to check that the above horizon $t= \pm R$ with respect
to the observer static at the spatial coordinate origin satisfies
this equation. What is called cosmological event horizons in
\cite{GH77} are now seen to correspond to `3-planes' tangent to the
`absolute' $\sigma(x)=0$, where the `absolute' and `3-planes' are in
terminology of projective geometry.

\begin{figure}
\centerline{\includegraphics[scale=1]{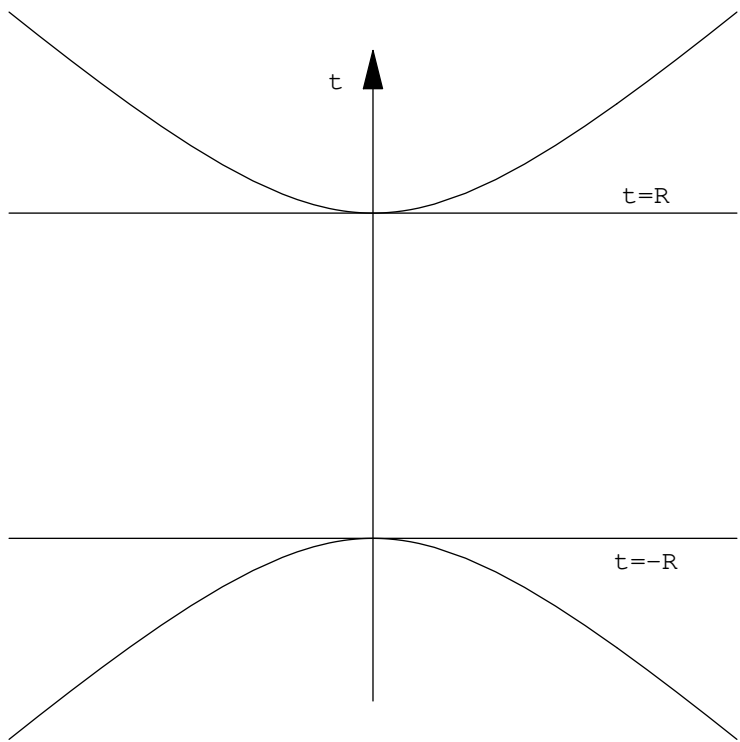} }
\caption{One patch of Beltrami coordinate system for de Sitter spacetime.
The vertical axis is
Beltrami time.  The region between two hyperbolae is the de Sitter
spacetime.  The two hyperbolae are projection boundary (and are also the conformal boundary) of de Sitter spacetime, where $\sigma(x)=0$.
The two horizontal lines are the event horizons for
the observer at spatial origin.  Each observer with $x^i=const.$ is
represented by vertical straight line, parallel to the Beltrami time axis.}
\end{figure}


At the horizon $|t|=R$ in Fig. 1, the values of $x^i$ can be taken
arbitrarily. In the \dS-spacetime in Beltrami coordinates, the world lines $x^i=const$ are all straight
lines.  A `test' mass along any of them  has vanishing
{\it coordinate} velocity.
Its 4-acceleration and 3-(coordinate) acceleration  are
both zero. Thus,  the {\it coordinate acceleration} at horizon is
\begin{eqnarray}\label{a=0}%
 \lim_{|t|\to R}\frac {\upd^2 x^\mu}{\upd t^2}=0 %
 \end{eqnarray}
for `test' mass at $x^\alpha=const$.  In fact,  the inertial
motion (\ref{imotion}) is equivalent to %
\be\label{tgeodesic}%
 m \frac {\upd}{\upd s} \left ( {\si^{-1} (x)} \frac {\upd
x^\mu}{\upd s}\right ) = 0 , \ee%
which is the time-like geodesic, and  the second law of mechanics
can
be written as \cite{SRBdSc}%
 \be m\frac {\upd}{\upd s} \left ( {\si^{-1} (x)} \frac {\upd x^\mu}{\upd s}\right ) = F^\mu ,%
 \ee%
  where
$F^\mu$ is the 4-force pseudo-vector. Thus, the equivalence of eq.
(\ref{a=0}) to eq. (\ref{tgeodesic}) 
means that there is no force needed for an inertial observer to
hold a `test' mass in place. In fact, the deviation of two nearby
inertial observers who are static to each other keeps a constant. In
other words, the relative 3-acceleration is also zero. Therefore,
there is indeed no `surface gravity' on the event horizon in
\dS-spacetime with a Beltrami metric.

This is also true for the horizon given by eq.~(\ref{Heq}) for other
{\it inertial} observers. The reason is the same as the one at
$|t|=R$. In fact, one may find a mapping corresponding to an $FLT$
sending $\tilde A(\tilde a^i)$ to $A(a^i|_{a^i=0})$, which maps a
point $\tilde A'( \tilde a')\in \mathscr{I}^++\mathscr{I}^-$ to its
counterpart $A'(a') \in \mathscr{I}^++\mathscr{I}^-$.  Thus, in view of
de Sitter invariant special relativity, all such a kind inertial
observers move with uniform coordinate 3-velocities along straight
lines.  Thus, the horizon on \dS-spacetime with Beltrami metric should be without `surface
gravity'.

On the other hand, it can be shown that the non-vanishing
surface gravity on \dS-horizon given in \GR\ is actually a kind
of {\it inertial force}, which leads to the departure from a
uniform-velocity motion along straight line.

By the Bekenstein-Hawking relation, this leads to the conclusion that
Hawking temperature in \dS-spacetime in view of inertial observers in de Sitter invariant special relativity
should be zero.  This is consistent with the previous result from the
argument of the periodicity in the imaginary time.

\section{Conclusions}\label{conclusion}

Different from the approach in \GR, which leads to some of the
\dS-puzzles, in view of an inertial observer in \dS-spacetime, the
`surface gravity' on the horizon in Beltrami coordinates should
vanish. And the Beltrami-time coordinate has no imaginary period.
These imply that field theories defined in Beltrami coordinate
systems with $SO(4,1)$ invariance should be at
zero-temperature in analog with the ones in Minkowski-spacetime. %
In the \dS-spacetime with a
static metric, the non-vanishing surface gravity should
also be regarded as a kind of inertial force and the `cosmic time'
coordinate has an imaginary period. They mean that the field
theories defined on it should be finite-temperature and the entropy
appears. The temperature is $(2\pi R)^{-1}$.

This is also supported by the relation between a Beltrami metric and
static metric of \dS-spacetime that is in analog with the relation between Minkowski
metric and Rindler metric in flat spacetime.

Thus, it seems that de Sitter spacetime can be understood from a
dramatically different view from that in \GR.  In the new view, the
Beltrami coordinates and metric have very special and important
meaning for the \dS-spacetime.  They should be regarded as
the most fundamental ones than all others and the observables should
be defined on the Beltrami systems as the inertial coordinate
systems in \dS-spacetime. Although there exists a horizon it is
not needed to search for the statistical origin of the entropy,
which in fact should be an irrelevant concept, for such a
horizon, since it is at zero temperature. And the entropy of the
horizon in the \dS-spacetime with a static metric is analogous to
the entropy of the Rindler horizon, caused by non-inertial motion and non-inertial
coordinates in view of {\it inertial observers in \dS-spacetime.}

It is worth to explore how far we can go in
the new viewpoint.  It is also worth to explain why the different
approaches in \dS-spacetime with a Beltrami metric give rise to the same conclusion and to
determine whether the conclusion is valid only in Beltrami
coordinate system.

\acknowledgments
The authors would like to thank Professors Z. Chang,
 G.W. Gibbons, Q.K. Lu and Z. Xu as well as Dr.  Y. Tian %
for valuable discussions and comments. This work is partly
supported by NSFC under Grants Nos. 90103004, 90403023, 10375087, 10373003, 10505004.

\end{document}